\RequirePackage{fix-cm}

\documentclass[smallextended]{svjour3}       % onecolumn (second format)
\smartqed  % flush right qed marks, e.g. at end of proof

\usepackage{graphicx}
\usepackage{amsfonts}
\usepackage{color}
\usepackage{soul}
\journalname{Quantum Studies: Mathematics and Foundations}

\begin{document}

\title{A Family of Weyl-Wigner Transforms for Discrete Variables Defined in a Finite-Dimensional Hilbert Space
}


\titlerunning{Family of Weyl-Wigner Transforms for Discrete Variables}        % if too long for running head

\author{Ady Mann         
\and 
Pier A. Mello 
\and
Michael Revzen
%etc.
}

\authorrunning{Mann, Mello, Revzen} % if too long for running head

\institute{A. Mann \at
       Department  of Physics, Technion-Israel Institute of Technology, Haifa 32000, Israel     \\
              \email{ady@physics.technion.ac.il}           %  \\
           \and
           P. A. Mello \at
              Instituto de F\'isica, Universidad Naciional Aut\'onoma de M\'exico, 
              Apartado Postal 20-364, M\'exico D. F., 01000 Mexico \\
              \email{mello@fisica.unam.mx}
            \and
            M. Revzen  \at 
            Department  of Physics, Technion-Israel Institute of Technology, Haifa 32000, Israel                     \\
         \email{revzen@physics.technion.ac.il}
}

\date{Received: date / Accepted: date}

\maketitle
%\hspace{5mm}
%{\bf Version: *mann-mello-revzen-resub-3*,   
%\hspace{2mm}
%Fri., Nov. 4, 2016, 8pm}
%\marginpar{{\bf REMOVE!!}}
%\vspace{5mm}

%%%%%%%%%%%%%%%%%%%%
\begin{abstract}
We study the Weyl-Wigner transform in the case of discrete variables defined in a Hilbert space of finite prime-number dimensionality $N$.
We define a family of Weyl-Wigner transforms as function of a phase parameter. 
We show that it is only for a specific value of the parameter that all the properties we have examined have a parallel with the case of continuous variables defined in an infinite-dimensional Hilbert space.
A geometrical interpretation is briefly discussed.
%%%%%%%%%%%%%%%%%%%%%%%%%%%%%%%%%%%

\end{abstract}
%%%%%%%%%%%%%%%%%%%%

%%%%%%%%%%%%%%%%%%%%
\section{Introduction}
\label{intro}

The Weyl-Wigner transform (WWT) was originally introduced
to provide a phase-space representation of an operator defined in a continuous Hilbert space \cite{weyl,wigner,hillery_et_al_1984,moyal49}
(see also \cite{leonhardt,schleich}).
When the operator in question is the density operator $\hat{\rho}$, we speak of the Wigner function (WF) of the state.

For a discrete Hilbert space of finite dimensionality, the WWT has been treated extensively in the literature. A representative list of references is given by
\cite{buot74,hannay-berry80,cohen-scully86,wootters87,galetti-de-toledo-piza88,cohendet-et-al88,kasperkovitz-peev94,leonhardt95-96,luis-perina98,rivas-ozorio99,bandyopadhyay_et_al_2001_2002,gibbons-et-al04,vourdas04,klimov_et_al_2006,revzen-WF012,mello-revzen-PRA2014,klimov_et_al_2016};
see also the references contained therein.
In Ref. \cite{mello-revzen-PRA2014} a definition was proposed which was appropriate for establishing a relation of WF with Kirkwood's joint quasi-probability distribution and with the von Neumann model of measurement.

Various authors 
(see, e.g., Refs.\cite{wootters87,klimov_et_al_2006,klimov_et_al_2016}) 
have found a certain freedom in the election of the phase in the definition of the (finite dimensional) Wigner function. 
%To the best of our understanding, the freedom contemplated in the present %paper is different in nature.
In the present paper we show that the definition for the discrete case 
given in Ref. \cite{mello-revzen-PRA2014} can also be extended
to define a family of WWTs as function of a phase parameter $c$.
We study this freedom in detail and show that one particular case, $c=0$, reproduces the definition given in Ref. \cite{mello-revzen-PRA2014},
while another member of the family, $c=-1/2$, brings the various expressions for the discrete case to a form similar to that for the continuous case.

The paper is organized as follows.
In Sec. \ref{discrete WT} we define the family of WWTs for the discrete case, describe some of its properties, and analyze the 
special role played by the choice $c=-1/2$ in the structure of the discrete WWT when it is compared with the continuous case.
The approach we adopt in this presentation provides a somewhat different view of the intimate role (discovered by Grossmann and Royer \cite{grossmann,royer}) of parity in the phase-space formulation of QM.
In Sec. \ref{WWT cont variables} we present an alternative form of the WWT for the continuous case, which exhibits its relation to the discrete one from another angle.
In Sec. \ref{geometry} we briefly discuss a geometrical interpretation
of our approach and compare it with that studied by other authors.
Sec. \ref{concl} contains some concluding remarks.
For coherence and completeness, and in order not to interrupt the main flow of the presentation, we relegate to the appendices some proofs and discussions of properties of various concepts used in the text.

%%%%%%%%%%%%%%%%%%%%%%%%%%%
\section{Discrete Weyl-Wigner Transform}
\label{discrete WT}

\subsection{Definition and properties}
\label{defin WT discrete}

The possibility of defining WWT for a Hilbert space of finite dimensionality
has been studied by many authors
\cite{buot74,hannay-berry80,cohen-scully86,wootters87,galetti-de-toledo-piza88,cohendet-et-al88,kasperkovitz-peev94,leonhardt95-96,luis-perina98,rivas-ozorio99,bandyopadhyay_et_al_2001_2002,gibbons-et-al04,vourdas04,klimov_et_al_2006,revzen-WF012,mello-revzen-PRA2014,klimov_et_al_2016}.
Here we extend the definition of Ref. \cite{mello-revzen-PRA2014} and 
define a family of WWTs:
using the Schwinger operators $\hat{Z}$ and $\hat{X}$
(see Ref. \cite{schwinger}, where these operators are designated as $U$ and $V$)
for a Hilbert space of a prime-number dimensionality $N$ (summarized in App. \ref{schwinger})
we define, for an operator $\hat{A}$, 
and as function of the parameter $c$, the family
%\begin{widetext}
%%%%%%%%%%%%%%%%%%%%%%
\begin{eqnarray}
W_{\hat{A}}^{(c)}(q,p)
=  \frac{1}{N}
\left\{
{\rm Tr}\sum_{b=0}^{N-1}\sum_{k=1}^{N-1}
\hat{A}\left[\left(\hat{X}\hat{Z}^b \right)^k \right]^{\dagger}
e^{i\frac{2\pi}{N}k[-p + b (q+c)]}  \right. &&
\nonumber \\
 \hspace{6cm} \left. + {\rm Tr}\sum_{k=0}^{N-1}
\hat{A} \left(\hat{Z}^k\right)^{\dagger}  e^{i\frac{2\pi}{N}k q}
 \right\}\; .  
\label{discreteWF 1}
\end{eqnarray}
%%%%%%%%%%%%%%%%%%%%%%
%\end{widetext}
%Recalling Eqs. (\ref{X(p)}), (\ref{Z(q)}), we notice that Eq. (\ref{discreteWF 1}) %defining WT for the discrete case is closer in structure to Eq. (\ref{WF continuous %3}) than to Eq. (\ref{WF continuous a})
%for the continuous case.
%We shall discuss this relation further in Sec. \ref{relation discrete-continuous} %below.
%%The phase $c$ in the above definition introduces a family of WT's;
%while the choice $c=0$ reproduces the definition given in Ref. \cite{mello-revzen-%PRA2014},
%We shall find below the particular choice that brings the various expressions for %the discrete case close to those for the continuous case discussed in the previous %%section.
We have defined a {\em discrete phase space} in which the 
``coordinate-like" and ``momentum-like" variables
are denoted by $q,p = 0,1, \cdots, N-1$.
%which thus consists of an $N \times N$ set of points.
The  operators $\left(\hat{X}\hat{Z}^b \right)^k$
[$N(N-1)$ in number: $b=0,1, \cdots, N-1; \; k=1, \cdots, N-1$], together with the $N$ operators $\hat{Z}^k$
($k=0, \cdots, N-1$) which appear in Eq. (\ref{discreteWF 1}), form a complete set of $N^2$ operators.
Notice that this set of operators is the finite-dimensional analog of the continuous complete set $\exp[i(u\hat{q}+v\hat{p})]$.
If $N$ is a prime number, the integers $0,1, \dots, N-1$ form an algebraic field analogous
to that of the real numbers in the continuous case.
The $N=2$ case requires a special treatment.
We note that the exponents of $\omega = {\rm exp}(2\pi i/N)$, which will appear frequently in our analysis, always belong to the  ${\rm Mod}[N]$ algebra. 
We remark that, when $N$ is a prime number, 
there is a simple expression for the
%the problem admits exactly 
$N+1$ {\em mutually unbiased bases} (MUB) \cite{ivanovic}, as summarized in App. \ref{mub}
(see, e.g., Refs. 
\cite{bandyopadhyay_et_al_2001_2002,revzen-WF012,durt_et_al}).

The definition (\ref{discreteWF 1}) can be given an alternative expression,
whose interest lies in the fact that it makes its manipulation easier and, also, suggests an interesting geometrical interpretation:
the expression (\ref{discreteWF 1}) can be written in terms of MUB as
%%%%%%%%%%%%%%%%%%%%%%
\begin{eqnarray}
W_{\hat{A}}^{(c)}(q,p)
&=&\frac{1}{N}
\sum_{b=\ddot{0}}^{N-1} \sum_{k=0}^{N-1} \sum_{m=0}^{N-1}
e^{\frac{2\pi i}{N}k \big[M_{q,p}^{(c)}(b)-m \big]}
\left\langle m; b\left|\hat{A}\right|m;b\right\rangle
- {\rm Tr} (\hat{A})  \; ,
\label{discreteWF 2 a}
\end{eqnarray}
%%%%%%%%%%%%%%%%%%%%%%
where the reference basis (eigenstates of $Z$) has been denoted, for convenience, as $\ddot{0}$,
and we have defined
%%%%%%%%%%%%%%%%%%%%%%
\begin{equation}
M_{q,p}^{(c)}(b) =
\left\{
\begin{array}{cl}
[-p+b(q+c)] \; {\rm Mod}[N],  & {\rm for} \;\;\; b = 0, \cdots, N-1 \; , \\
q, & {\rm for} \;\;\; b=\ddot{0} \; .
\end{array}
\right.
\label{m(b)}
\end{equation}
%%%%%%%%%%%%%%%%%%%%%%
%For a given pair of variables $q,p$, Eq. (\ref{m(b)}) states that, for $b=\ddot{0}$, %$M_{q,p}(\ddot{0})=q$;
%for $b=0$, $M_{q,p}(0)=-p \; {\rm Mod}[N]=N-p$;
%for subsequent values of $b$,
%$M_{q,p}(b)=(-p+b(q+c)) \; {\rm mod}[N]$.
%Thus, $M_{q,p}(b)$ may be viewed as specifying ``points" in a $b-m$ plane:
%$b$ is along the $x$-axis and takes the values $b=\ddot{0},0,1,\cdots,N-1$, which denote %the $N+1$ bases;
%$m$ is along the $y$-axis and takes the values $b=0,1,\cdots,N-1$, which denote the $N$ %states for each basis.
%This aggregate of points, for {\em fixed} $q$ and $p$, may be described as a ``line" in %the $b-m$ plane.
%We thus refer to $M_{q,p}(b)$ as a line, and its corresponding operator, $\hat{P}_{q,p}$, %Eq. (\ref{Pj}) below, as a line operator;
%it is similar to the ``phase-point" operator introduced in Secs. V and VI of
%Ref. \cite{wootters87}.

In Eq. (\ref{discreteWF 2 a}) we can do the sum over $k$,
%using the result
%%%%%%%%%%%%%%%%%%%%%%
%\begin{equation}
%\frac{1}{N}\sum_{k=0}^{N-1}
%e^{\frac{2\pi i}{N}k \big[M_{q,p}(b)-m \big]}
%= \delta_{m,\; M_{q,p}(b)} \; ,
%\label{sum_k}
%\end{equation}
%%%%%%%%%%%%%%%%%%%%%%
%where the arguments of the Kronecker delta are understood to be ${\rm Mod}[N]$.
%in other words, for given $q,p$, the sum (\ref{sum_k}) vanishes unless $m$ equals
%$[-p+b(q+c)] \; {\rm Mod}[N]$ when
%$b \neq \ddot{0}$, or $q$ when $b = \ddot{0}$.
so Eq. (\ref{discreteWF 1}) can be given the alternative form
%(for details, see Ref. \cite{mello-revzen-PRA2014})
%%%%%%%%%%%%%%%%%%%%%%
%\numparts
%\begin{subequations}
\begin{eqnarray}
W_{\hat{A}}^{(c)}(q,p)
&=& \sum_{b=\ddot{0}}^{N-1}
\left\langle M_{q,p}^{(c)}(b);b\left|\hat{A}\right|M_{q,p}^{(c)}(b);b \right\rangle
%\nonumber \\
%&& \hspace{1cm} 
- {\rm Tr} (\hat{A})  \; ,
\label{discreteWF 3 a} \\
&=& {\rm Tr} (\hat{A}\hat{P}^{(c)}_{q,p}) \; ,
\label{discreteWF 3 b}
\end{eqnarray}
%\end{subequations}
%%%%%%%%%%%%%%%%%%%%%%
where we have defined the Hermitean operator
%%%%%%%%%%%%%%%%%%%%%%
\begin{eqnarray}
\hat{P}^{(c)}_{q,p}
=\sum_{b=\ddot{0}}^{N-1}
\big|M_{q,p}^{(c)}(b);b\big\rangle  \big\langle M_{q,p}^{(c)}(b); b \big|
-\hat{\mathbb{I}},
\label{P(q,p)}
\end{eqnarray}
%\label{discrete_WF and line_operator}
%\endnumparts
%%%%%%%%%%%%%%%%%%%%%%
($\hat{\mathbb{I}}$ being the unit operator),
which we now analyze.
For a given pair of phase-space variables $q,p$, the expression $M^{(c)}_{q,p}(b)$ may be considered as defining ``points" in a $b$-$m$ plane: 
$b$ is along the horizontal axis and takes the values $b= \ddot{0}, 0, 1, \cdots, N-1$, which denote the $N+1$ bases;
$m$ is along the vertical axis and takes the values $m = 0, 1, \cdots, N-1$, which denote the $N$ states for each basis.
Thus, for fixed $q,p$, this set of points may be taken as defining a ``line" in the $b$-$m$ plane (see Fig. \ref{line_bm_plane} for an example):
we thus refer to the set $M^{(c)}_{q,p}(b)$ ($b=\ddot{0},0,\cdots,N-1$) as a line, and to the operator $\hat{P}^{(c)}_{q,p}$ of 
Eq. (\ref{P(q,p)}) as a {\em line operator}.

Our first aim is to obtain the matrix elements of the line operator.  
We split $\hat{P}^{(c)}_{q,p}$ as 
%%%%%%%%%%%%%%%%%%%%%%
\begin{eqnarray}
\hat{P}^{(c)}_{q,p} = (\hat{P}^{(c)}_{q,p})' + \hat{P}''_{q,p} \; ,
\label{P=P'+P''}   
\end{eqnarray}
%%%%%%%%%%%%%%%%%
where
%%%%%%%%%%%%%%%%%%%%%%
\begin{eqnarray}
(\hat{P}^{(c)}_{q,p})'
&=& \sum_{b=0}^{N-1}
\big| M^{(c)}_{q,p}(b);b\big\rangle  \big\langle M^{(c)}_{q,p}(b); b \big|
-\hat{\mathbb{I}} \; ,   
\label{P'}   \\
\hat{P}''_{q}
&=& |  q \rangle  \langle q  | ,
\hspace{5mm}
b = \ddot{0} \;  .
\label{P''}
\end{eqnarray}
%%%%%%%%%%%%%%%%%
We refer to $(\hat{P}^{(c)}_{q,p})'$ as the ``amputated" line operator and consider it first. 

A {\em diagonal} matrix element with respect to the reference basis of any one of the summands in the first term of Eq. (\ref{P'}) equals $1/N$, according to the definition of MUB,
as can also be verified from the explicit form of the MUB states given in  
Eq. (\ref{mub e-vectors}).
Thus {\em the amputated line operator $(P^{(c)}_{q,p})'$ has zero diagonal elements}, i.e.,
%%%%%%%%%%%%%%%%%
\begin{equation}
\langle n | (\hat{P}^{(c)}_{qp})' | n\rangle = 0 \; ,
\label{nP'n}
\end{equation}
%%%%%%%%%%%%%
because of the subtraction of $\mathbb{I}$. 

We now consider a {\em non-diagonal} matrix element for an arbitrary $b$.
From Eq. (\ref{mub e-vectors}) we find
%%%%%%%%%%%%%%%%%
\begin{equation}
\langle n | M^{(c)}_{q,p}(b);b\rangle \langle b;M^{(c)}_{q,p}(b)|n' \rangle
=\frac{1}{N}\omega^{(n-n')[\frac{b}{2}(n+n'-1)-M^{(c)}_{q,p}(b)]}.
\label{nn' m.el. of A_mb}
\end{equation}
%%%%%%%%%%%%%%%%%
If we compare this matrix element at $b$ with one at $b'\ne b$, we see that the two will be equal iff
%%%%%%%%%%%%%%%%%
\begin{equation}
\frac{n+n'-1}{2}=(q+c) \; {\rm Mod}[N],
\label{coherence}
\end{equation}
%%%%%%%%%%%%%%%%%
a condition to be referred to as the {\em coherence requirement}.
When we add up terms like (\ref{nn' m.el. of A_mb}) for $b=0, \cdots, N-1$ to construct the $n\neq n'$ matrix element of (\ref{P'}), only matrix elements with 
$n+n'=2q+2c+1$ will add up coherently; 
other matrix elements will cancel each other as they run over the roots of unity. 
Substituting, in Eq. (\ref{nn' m.el. of A_mb}), $n+n'=2q+2c+1$ and 
$M^{(c)}_{q,p}(b) = -p+b(q+c)$ from Eq. (\ref{m(b)}), we obtain, for the matrix elements 
$\langle n | (P^{(c)}_{qp})' | n' \rangle $, $N$ equal terms, with the result
%%%%%%%%%%%%%%%%%
\begin{equation}
\langle n | (\hat{P}^{(c)}_{qp})' | n' \rangle 
= \delta_{n+n', 2q+2c+1} \; \omega^{(n-n')p}, \;\;\;\; n \neq n' .
%\label{nP'n' n\neq n'}
\end{equation}
%%%%%%%%%%%%%%%%%

For the matrix elements of $P''_{qp}$ we have
%%%%%%%%%%%%%%%%%
\begin{equation}
\langle n | \hat{P}''_{qp} | n' \rangle 
= \delta_{nq}\delta_{n'q} \; .
\label{nP''n'}
\end{equation}
%%%%%%%%%%%%%%%%%

The matrix elements of the full line operator are then
%%%%%%%%%%%%%%%%%
\begin{equation}
\langle n | \hat{P}^{(c)}_{qp} | n' \rangle 
=\left\{
\begin{array}{cr}
 \delta_{n+n', 2q+2c+1} \; \omega^{(n-n')p},   &  \;\;\; n \neq n' \; ,\\
\delta_{nq} \; ,  & \;\;\;  n=n' \; ,
\end{array}
\right.
\label{nPn'}
\end{equation}
%%%%%%%%%%%%%%%%%
which can also be written, $\forall n,n'$, as
%%%%%%%%%%%%%%%%%
\begin{eqnarray}
\langle n | \hat{P}^{(c)}_{q p} | n' \rangle
=\delta_{q n} \delta_{q n'}
-\delta_{n n'} \; \delta_{n, \; q+c+\frac12} 
+ \delta_{n + n', 2q+2c+1} \; e^{\frac{2\pi i}{N}p(n-n')}.
\label{nPn' 2}
\end{eqnarray}
%%%%%%%%%%%%%%%%%

The structure of the matrix 
$\langle n | \hat{P}^{(c)}_{q p} | n' \rangle$
of Eqs. (\ref{nPn'}), (\ref{nPn' 2}), 
is illustrated for $N=5$ in Tables \ref{P(n,n')c=-1/2} and \ref{P(n,n')c=0}, for the values 
$c=-1/2$ and $c=0$ for the phase parameter, respectively.
For $c=-1/2$, the non-zero matrix elements for fixed $q$ appear precisely on the various secondary diagonals;
notice that, for a given $q$, the diagonal matrix element 
$\delta_{nq}\delta_{n'q}$ lies on the same secondary diagonal as the off-diagonal matrix elements; 
this is not the case for $c\ne -1/2$.
This shows the particular role played by $c=-1/2$.
\textcolor{blue}{We thus see that the phase freedom represented by the parameter $c$ has important consequences in the structure of the matrix of the line operator.}
We remark that this phase freedom can only be there for the discrete case; 
in the continuous case this is not allowed, if we insist that the resulting function be continuous.
It is thus the value $c=-1/2$ for the phase parameter in the discrete case that shows a similarity with the continuous case.

%%%%%%%%%%%%%%%%%%%%%%%%%%%%%%%%%%%%%%%%%%%%%%%
\begin{table}[ht]%\scriptsize
%\caption{A summary of the structural similarities between measurements in QM and CM.}
%\centering
\begin{tabular}{| c || c | c | c | c | c || }
\hline
$\downarrow n$ / $n'\rightarrow$ & 0 & 1 & 2 & 3 & 4 \\
\hline\hline
$0$ & $\delta_{n0}\delta_{n'0}$ & $$ & \fbox{\hl{$q=1$}} & & $q=2$
\\
\hline
$1$ & $$ & \fbox{\hl{$\delta_{n1}\delta_{n'1}$}} &  & $q=2$ & 
\\
\hline
$2$ & \fbox{\hl{$q=1$}} & & $\delta_{n2}\delta_{n'2}$ & & $q=3$
\\
\hline
$3$ & & $q=2$ & & $\delta_{n3}\delta_{n'3}$ &
\\
\hline
$4$ & $q=2$ & & $q=3$ & & $\delta_{n4}\delta_{n'4}$
\\
\hline\hline
\end{tabular}
\caption{Structure of the matrix 
$\langle n | \hat{P}^{(c)}_{q p} | n' \rangle$
of Eqs. (\ref{nPn'}), (\ref{nPn' 2}), for $N=5$ and $c=-1/2$.
The non-zero matrix elements for fixed $q$
(which obey $n+n'=2q$) appear on the various secondary diagonals.
For a given $q$, the diagonal matrix element 
$\delta_{nq}\delta_{n'q}$ (i.e., $n=n'=q$) appears precisely on the same secondary diagonal as the off-diagonal matrix elements; 
this is emphasized for the particular case $q=1$, where the corresponding matrix elements have been highlighted and a box has been drawn around each one of them.
}
\label{P(n,n')c=-1/2}
\end{table}
%%%%%%%%%%%%%%%

%%%%%%%%%%%%%%%%%%%%%%%%%%%%%%%%%%%%%%%%%%%%%%%
\begin{table}[ht]%\scriptsize
%\caption{A summary of the structural similarities between measurements in QM and CM.}
%\centering
\begin{tabular}{| c || c | c | c | c | c || }
\hline
$\downarrow n$ / $n'\rightarrow$ & 0 & 1 & 2 & 3 & 4 \\
\hline\hline
$0$ & $\delta_{n0}\delta_{n'0}$ & $q=0$ & & \fbox{\hl{$q=1$}} & $q=4$
\\
\hline
$1$ & $q=0$ & \fbox{\hl{$\delta_{n1}\delta_{n'1}$}} &\fbox{\hl{$q=1$}}  & $q=4$ & $q=2$
\\
\hline
$2$ & & \fbox{\hl{$q=1$}} & $\delta_{n2}\delta_{n'2}$ & $q=2$ & 
\\
\hline
$3$ & \fbox{\hl{$q=1$}} & $q=4$ & $q=2$ & $\delta_{n3}\delta_{n'3}$ &$q=3$
\\
\hline
$4$ & $q=4$ & $q=2$ & & $q=3$ & $\delta_{n4}\delta_{n'4}$
\\
\hline\hline
\end{tabular}
\caption{Structure of the matrix 
$\langle n | \hat{P}^{(c)}_{q p} | n' \rangle$
of Eqs. (\ref{nPn'}), (\ref{nPn' 2}), for $N=5$ and $c=0$.
The non-zero, off-diagonal matrix elements for fixed $q$ 
(which obey $n+n'=2q+1$) appear on the various secondary diagonals.
For a given $q$, the diagonal matrix element 
$\delta_{nq}\delta_{n'q}$
{\color{blue}(i.e., $n=n'=q$)} does {\em not} lie on the same secondary diagonal as the off-diagonal matrix elements; 
this is emphasized for the particular case $q=1$, where the various matrix elements have been highlighted and and a box has been drawn around each one of them.
}
\label{P(n,n')c=0}
\end{table}
%%%%%%%%%%%%%%%

An important property of the operators $\hat{P}^{(c)}_{q,p}$ is that they form a complete orthogonal set of $N^2$ operators.
For any value of $c$:

i) They fulfill the orthogonality relation
%%%%%%%%%%%%%%%%%%%%%%
%\numparts
\begin{eqnarray}
\frac{1}{N} {\rm Tr}\left[ \hat{P}^{(c)}_{q,p} \; \hat{P}^{(c)}_{q',p'} \right]
= \delta_{q, q'}\delta_{p, p'} \; .
\label{orthogon. Pj}
\end{eqnarray}
%%%%%%%%%%%%%%%%%%%%%%
This statement is proved in App. \ref{orthogonality of P's} employing geometrical arguments. 

ii) They satisfy the closure relation
%%%%%%%%%%%%%%%%%%%%%%
\begin{eqnarray}
\frac{1}{N} \sum_{q,p=0}^{N-1} \hat{P}^{(c)}_{q,p} = \mathbb{I} \; .
\label{closure discrete}
\end{eqnarray}
%\label{orthog-closure discrete}
%\endnumparts
%%%%%%%%%%%%%%%%%%%%%%
This follows from  Eq. (\ref{nPn'}): 
for $n \neq n'$, the sum over $p$ gives zero;
for $n=n'$ we have $\sum_{qp}\delta_{nq}=N$.

iii) An $N \times N$ matrix $\hat{A}$ can thus be written as a linear combination of the 
$\hat{P}^{(c)}_{q,p}$'s, i.e.,
%%%%%%%%%%%%%%%%%%%%%%
%\numparts
\begin{eqnarray}
\hat{A}
&=& \frac{1}{N} \sum_{q,p=0}^{N-1}
{\rm Tr}\left(\hat{A} \hat{P}^{(c)}_{q,p}\right)\hat{P}^{(c)}_{q,p}
\label{A as lc of Pj a}   \\
&=& \frac{1}{N} \sum_{q,p=0}^{N-1}
 W^{(c)}_{\hat{A}}(q,p) \hat{P}^{(c)}_{q,p} \; ,
\label{A as lc of Pj b}
\end{eqnarray}
%%%%%%%%%%%%%%%%%%%%%%
where we have used Eq. (\ref{discreteWF 3 b}).

From the above discussion we find that the WWT of Eqs. (\ref{discreteWF 1}) and
(\ref{discreteWF 3 b}) possesses the following properties.

1) For $\hat{A}=\hat{\rho}$, the WF is normalized as
%%%%%%%%%%%%%%%%%%%%%%
\begin{equation}
\frac{1}{N}
\sum_{p,q=0}^{N-1} W^{(c)}_{\hat{\rho}}(q,p)
=1 \; .
\label{normal}
\end{equation}
%%%%%%%%%%%%%%%%%%%%%%

2) The WWT of a Hermitean operator $\hat{A}$ is real, i.e.,
%%%%%%%%%%%%%%%%%%%%%%
\begin{equation}
W^{(c)}_{\hat{A}}(q,p) = [W^{(c)}_{\hat{A}}(q,p)]^{\star} \;, \;\;\;{\rm for}\;\;\; A^{\dagger}=A.
\label{W=W*}
\end{equation}
%%%%%%%%%%%%%%%%%%%%%%
This follows immediately from the Hermiticity of the operators $\hat{P}^{(c)}_{q,p}$.

3) The WWT of the unit operator is 1, i.e.,
%%%%%%%%%%%%%%%%%%%%%%%%%
\begin{equation}
W^{(c)}_{\hat I}(q,p) = 1 \; .
\label{WI = 1 discrete}
\end{equation}
%%%%%%%%%%%%%%%%%%%%%%%%

4) The WWTs of the operators $\hat{A}$ and $\hat{B}$ fulfill the so-called ``product formula" (see also Ref. \cite{wootters87}, Eq. (15))
%%%%%%%%%%%%%%%%%%%%%%
\begin{equation}
\frac{1}{N}
\sum_{q,p=0}^{N-1} W^{(c)}_{\hat{A}}(q,p) W^{(c)}_{\hat{B}}(q,p)
= {\rm Tr}(\hat{A} \hat{B}) \; .
\end{equation}
%%%%%%%%%%%%%%%%%%%%%%

%i) 5) From Eq. (\ref{discreteWF 3 b}), the WT of the projector
%$\hat{\mathbb{P}}_{mb}=|m,b\rangle \langle m,b|$ is
%%%%%%%%%%%%%%%%%%%%%%
%\begin{equation}
%W_{\hat{\mathbb{P}}_{mb}}(q,p)
%= {\rm Tr} (\hat{\mathbb{P}}_{mb} \hat{P}_{qp})
%= \delta_{M_{q,p}(b),m}  .
%\label{WT Pmb}
%\end{equation}
%%%%%%%%%%%%%%%%%%%%%%

5) The WF $W^{(c)}_{\hat{\rho}}(q,p)$ satisfies the marginality property, written in terms of the projector 
$\hat{\mathbb{P}}_{mb}= |mb\rangle \langle mb|$,
%%%%%%%%%%%%%%%%%%%%%%
\begin{equation}
{\rm Tr}\left(\hat{\rho} \; \hat{\mathbb{P}}_{mb} \right)
=\langle m,b| \hat{\rho} |m,b\rangle
= \frac{1}{N}\sum_{q,p=0}^{N-1}W^{(c)}_{\hat{\rho}}(q,p) \delta_{M_{q,p}^{(c)}(b),m} \; ,
\label{marginality disc}
\end{equation}
%%%%%%%%%%%%%%%%%%%%%%
where we recall that $M^{(c)}_{q,p}(b)$ is defined in Eq. (\ref{m(b)}).

Eq. (\ref{marginality disc}) 
%is analogous to (\ref{marginality cont}) for the continuous %case.
states that the probability to find the system
in the state $m$ of the basis $b$ (of our set of $N+1$ MUBs) is $1/N$ times the sum of the WF over %the points in the phase-space plane $q,p$ that satisfy
$M^{(c)}_{q,p}(b)=m$.
Its RHS
%of Eq. (\ref{marginality disc})
can be considered as defining the {\em Radon transform}
of the WF $W^{(c)}_{\hat{\rho}}(q,p)$
(see, e.g., 
Refs. \cite{leonhardt,schleich,revzen-WF012,khanna-mello-revzen}).
Two particular cases of the above marginality property are
%%%%%%%%%%%%%%%%%%%%%%
%\numparts
\begin{eqnarray}
\langle q_0| \hat{\rho} |q_0\rangle
&=& \frac{1}{N}\sum_{p}W^{(c)}_{\hat{\rho}}(q_0,p) \; ,
\label{marginal q0} \\
\langle p_0| \hat{\rho} |p_0\rangle
&=&\frac{1}{N}\sum_{q}W^{(c)}_{\hat{\rho}}(q,p_0) \; ,
\end{eqnarray}
%\endnumparts
%%%%%%%%%%%%%%%%%%%%%%
which are the standard marginality relations.

%%%%%%%%%%%%%%%%%%%%
\subsection{The particular case $c=-1/2$}
\label{c=-1/2}

We aready remarked that the choice $c=-1/2$ for the phase introduced in the original definition (\ref{discreteWF 1}) constitutes an important particular case of the general formalism.
We now examine this choice in relation with another property of the WF.

In this case, the matrix elements (\ref{nPn'}), (\ref{nPn' 2}) of the operator 
$\hat{P}^{(c)}_{q,p}$ reduce to the simpler form
(we remove, for brevity, the $c=-1/2$ index in the various expressions)
%%%%%%%%%%%%%%%%%
\begin{equation}
\langle n | \hat{P}_{qp} | n' \rangle 
= \delta _{n+n',2q}
\omega^{(n-n')p}, \;\;\;\; \forall n,n'.
\label{nPn' c=-1/2}
\end{equation}
%%%%%%%%%%%%%%%%%
For this choice, the operator $\hat{P}_{q,p}$ can be related to the parity operator 
$\hat{\Pi}_{q,p}$ around the point $(q,p)$ as 
\cite{grossmann,royer,zak2011}
%%%%%%%%%%%%%%%%%
\begin{equation}
\hat{P}_{q,p}= \hat{\Pi}_{q,p} \; .
\label{P(q,p) vs Pi discr}
\end{equation}
%%%%%%%%%%%%%%%%%
This is quite evident, e.g., when $q=p=0$, from Eq. (\ref{nPn' c=-1/2}).
Indeed,
%%%%%%%%%%%%%%%
\begin{eqnarray}
\langle n|\hat{P}_{0,0}|n' \rangle
&=&\delta_{n+n',0} 
\label{P00 m.els.}  \\
{\rm and \; hence} \hspace{1cm}
\hat{P}_{0,0}
&=&\sum_{n=0}^{N-1}|n\rangle\langle -n| \; ,
\label{P00}  
\end{eqnarray}
%%%%%%%%%%%%%%%%%
which is the parity operator around the point $(q,p)=(0,0)$.
More generally, the operator
%%%%%%%%%%%%%%%%%
\begin{equation}
\hat{\Pi}_{q,p}
=\hat{X}^q\hat{Z}^{p}\hat{P}_{0,0}\hat{Z}^{-p}\hat{X}^{-q},
\label{P(qp)}
\end{equation}
%%%%%%%%%%%%%%%%%
which can be seen to have the same $n,n'$ matrix elements as $\hat{P}_{q,p}$,
can be verified to correspond to the parity operator $\hat{\Pi}_{q,p}$ around the point $(q,p)$.
One can also easily verify that 
$\big(\hat{\hat{\Pi}}_{q,p}\big)^2=\mathbb{I}$, further verifying its shifted parity character.

This property, which we showed for $c = -1/2$, has a parallel in the WWT for the case of a Hilbert space for continuous variables
\cite{grossmann,royer,zak2011}.

The discussion given in the next section gives a better understanding why the choice $c=-1/2$ in the family of discrete WTTs allows us to achieve, in a natural way, a structure for the WWT which is similar to that for the continuous case.
%%%%%%%%%%%%%%%%%%%%%%%

%%%%%%%%%%%%%%%%%%%%%%
\section{Relation with the WWT for continuous variables}
\label{WWT cont variables}

We first re-write the definition (\ref{discreteWF 1}) of the WWT
for {\em discrete variables} in terms of 
position-like and momentum-like operators using (\ref{X(p)}) and (\ref{Z(q)}), as
%\begin{widetext}
%%%%%%%%%%%%%%%%%%%%%%
%\numparts
\begin{eqnarray}
W_{\hat{A}}(q,p)
%\nonumber \\
&=&\frac{1}{N}
\left\{
{\rm Tr}\sum_{b=0}^{N-1}\sum_{k=1}^{N-1}
\hat{A}\left[\left({\rm e}^{-\frac{2\pi i}{N}\hat{p}}
{\rm e}^{\frac{2\pi i}{N}b \hat{q}}\right)^k \right]^{\dagger}
e^{i\frac{2\pi}{N}k[-p + b (q+c)]}
\right.
\nonumber \\
&& \left. \hspace{4cm} + {\rm Tr}\sum_{k=0}^{N-1}
\hat{A} \left( {\rm e}^{\frac{2\pi i}{N}k \hat{q}} \right)^{\dagger}
e^{i\frac{2\pi}{N}k q }
\right\} ,
\label{discreteWF 10 a}
%\nonumber
 \\
&=&\frac{1}{N}
\left\{
{\rm Tr}\sum_{b=0}^{N-1}\sum_{k=1}^{N-1}
\hat{A}
\left({\rm e}^{-\frac{2\pi i}{N}\hat{p}}
{\rm e}^{\frac{2\pi i}{N}b \hat{q}}\right)^{-k}
e^{i\frac{2\pi}{N}k[-p + b (q+c)]}
\right.
\nonumber \\
&& \left. \hspace{4cm}
+ {\rm Tr}\sum_{k=0}^{N-1}
\hat{A} \;  {\rm e}^{-\frac{2\pi i}{N} k \hat{q}}
e^{i\frac{2\pi}{N}k q}
\right\} \; .
%\nonumber
\label{discreteWF 10 b}
%\tilde{W}_{\hat{A}}(k,b)
%&=& {\rm Tr}\left\{\hat{A}\left[\left(\hat{X}\hat{Z}^b \right)^k \right]^{\dagger} %\right\}     \; ,
%\label{discreteWF b}\\
%\tilde{W}_{\hat{A}}(l)
%&=& {\rm Tr}\left[\hat{A} \left(\hat{Z}^l\right)^{\dagger} \right] \; .
%\label{discreteWF c}
\end{eqnarray}
\label{discreteWF 10}
%\endnumparts
%%%%%%%%%%%%%%%%%%%%%%
We shall use this expression below.

We now turn to the case of {\em continuous variables}. The WWT can be expressed as the inverse Fourier transform of the characteristic function of the operator as \cite{leonhardt,schleich,mello-revzen-PRA2014}
%%%%%%%%%%%%%%%%%
\begin{equation}
W_{\hat{A}}(q,p)
= \frac{1}{2\pi} \int_{-\infty}^{\infty} \int_{-\infty}^{\infty}
{\rm Tr} \left[\hat{A} {\rm e}^{-i(u\hat{q} + v\hat{p})} \right] {\rm e}^{i(uq + vp)} du dv ,
\label{WF continuous a}
%\\
%\tilde{W}_{\hat{A}}(u,v)
%&=& {\rm Tr} \left[\hat{A} e^{-i(u\hat{q} + v\hat{p})} \right]  .
\end{equation}
%%%%%%%%%%%%%%%%%%%%%%%%%%%%%%%
%The definition (\ref{WF continuous}) is equivalent to the standard one, %presented, for convenience, in Eq. (\ref{WT,continuous,standard}).
where we have used units in which $q$ and $p$ are dimensionless, and $\hbar=1$.

It is shown in App. \ref{alternative structure for continuous case} that the definition 
(\ref{WF continuous a}) can be written in the alternative form
%%%%%%%%%%%%%
\begin{eqnarray}
W_{\hat{A}}(q,p)
&=& {\rm Tr}\int_{0}^{\infty} r dr
\int_{0}^{2 \pi} d \theta \;
\hat{A} 
\left( {\rm e}^{-i \hat{p}} \;{\rm e}^{i \cot \theta \; \hat{q}}\right)
^{-r \sin \theta}
\nonumber \\
&& \hspace{2cm}\times {\rm e}^{i r \sin \theta \left[- p + \cot \theta \; (q-\frac12)\right]} \; .
\label{WF continuous 3}
\end{eqnarray}
%%%%%%%%%%%%%
Notice that, having expressed the WWT in terms of the operator
${\rm e}^{-i \hat{p}} \;{\rm e}^{i \cot \theta \; \hat{q}}$ as in
Eq. (\ref{WF continuous 3}),
[instead of the operator ${\rm e}^{-i(v\hat{p} + u\hat{q})}$ used in
Eq. (\ref{WF continuous a})],
there appears naturally a shift $-1/2$ in the coordinate $q$.

We now compare Eq. (\ref{discreteWF 10 b}) for discrete variables with 
Eq. (\ref{WF continuous 3}) for the continuous case. 
Notice that the first term in Eq. (\ref{discreteWF 10 b}) has the structure of Eq. (\ref{WF continuous 3}) for the continuous case when 
$\theta \neq 0$,
if we make the correspondences shown in Table \ref{identific}.
For $\theta =0$, we notice that
%%%%%%%%%%%%%
\begin{eqnarray}
\lim_{\theta \to 0} 
\left\{
\left( {\rm e}^{-i \hat{p}} \;{\rm e}^{i \cot \theta \; \hat{q}}\right)
^{-r \sin \theta}
{\rm e}^{i r \sin \theta \left[- p + \cot \theta \; (q-\frac12)\right]}
\right\} = {\rm e}^{-ir\hat{q}}{\rm e}^{irq} ,
\end{eqnarray}
%%%%%%%%%%%%%
%we go back to the form (\ref{WF continuous 2 a}) for the Wigner Transform: the
%integrand reduces to ${\rm e}^{-ir\hat{q}}{\rm e}^{irq}$, 
which has the structure of the
second term of Eq. (\ref{discreteWF 10 b}), identifying $k$ with $r$.
%%%%%%%%%%%%%%%%%%%%%%%%%%%%%%%%%%%%%%%%%%%%%%%
\begin{table}[ht]%\scriptsize
%\caption{A summary of the structural similarities between measurements in QM and CM.}
%\centering
\begin{tabular}{| c
%p{3.4cm}
| c | %l |
}
\hline
Eq. (\ref{discreteWF 10 b}) & Eq. (\ref{WF continuous 3})  %&
\\
\hline\hline
${\rm e}^{-\frac{2\pi i}{N}\hat{p}}
{\rm e}^{\frac{2\pi i}{N}b \hat{q}}$
&
${\rm e}^{-i \hat{p}} \;{\rm e}^{i \cot \theta \; \hat{q}}$
%& Eq.~(\ref{<Q>'CM}): $\frac{\langle Q \rangle'}{\epsilon} = \langle A  \rangle$
\\
\hline
$b$
& $\cot \theta$
%& Eq.~(\ref{CM-uncert-q}): $\sim \frac{\sigma_Q}{\epsilon}$
\\
\hline
$k$
& $r \sin \theta$
%& Eq.~(\ref{rho'_s 1}): $\rho'_{s}(q,p)=  e^{\tau \hat{A}_{op}^2} \; \rho_s(q,p)$
\\
\hline
$c$
& $-1/2$
%& Eq.~(\ref{CM diffus_eqn a}): $\frac{\partial \rho'_{s}(q,p)}{\partial \tau}
%= \hat{A}_{op}^2 \rho'_{s}(q,p)$
\\
\hline\hline
\end{tabular}
\caption{Correspondence of the discrete and continuous WWT's, 
for $\theta \neq 0$.}
\label{identific}
\end{table}
%%%%%%%%%%%%%%%
%%%%%%%%%%%%%%%%%%%%%%%%%%%%%%%%%%%
%\begin{widetext}
%%%%%%%%%%%%%
%\numparts
%\begin{eqnarray}
%\hat{X}\hat{Z}^b=
%&&{\rm e}^{-\frac{2\pi i}{N}\hat{p}}
%{\rm e}^{\frac{2\pi i}{N}b \hat{q}}, \;\;
%{\rm Eq. \; (\ref{discreteWF 10 b})}, \;\;
%{\rm with} \;\;
%{\rm e}^{-i \hat{p}} \;{\rm e}^{i \cot \theta \; \hat{q}}, \;\;
%{\rm Eq.  \; (\ref{WF continuous 3})}
%\nonumber \label{identif-disc-cont a}   \\
%&&b, \hspace{2.5cm} '' \hspace{1.8cm}  {\rm with} \;\;\; \cot \theta ,
%\hspace{2cm}   ''
%\nonumber \label{identif-disc-cont b}  \\
%&&k, \hspace{2.5cm} '' \hspace{1.8cm}  {\rm with} \;\;\; r \sin \theta ,
%\hspace{1.8cm}   ''
%\nonumber \label{identif-disc-cont c}  \\
%&&c, \hspace{2.5cm} '' \hspace{1.8cm}  {\rm with} \;\;\; -1/2 ,
%\hspace{1.8cm}   ''
%\nonumber \label{identif-disc-cont d}
%\end{eqnarray}
%\label{identif-disc-cont}
%\endnumparts
%%%%%%%%%%%%%
%\end{widetext}
However, the fact that the $-1/2$ shift in $q$ goes away for $\theta =0$ is of no consequence in the continuous-variable case, as it occurs in a set of zero measure and thus does not affect the integral (36).
This is consistent with our previous discussion in relation with 
Tables 1 and 2, where we noted that the phase freedom that we have in the discrete case is not there in the continuous case.
 
The above discussion gives a better understanding why the choice $c=-1/2$ for the discrete case gives a WWT having similar properties as for the continuous case: in point of fact,
in the latter, continuous case, a shift $-1/2$ in $q$ occurs in a natural way when we express the operator 
${\rm e}^{-i(v\hat{p} + u\hat{q})}$ 
in terms of the product of operators
${\rm e}^{-i \hat{p}} \;{\rm e}^{i \cot \theta \; \hat{q}}$ 
so as to give it a similar structure as in the former, discrete case.
%%%%%%%%%%%%%%%%%%%%%%%

%%%%%%%%%%%%%%%%%%%%%%
\section{Geometrical interpretation of the ``phase freedom" and its relation with other approaches}
\label{geometry}

As we mentioned in Sec. \ref{discrete WT}, the considerations briefly mentioned there lead us naturally to a geometrical interpretation of our analysis (see, e.g., Ref. \cite{revzen-WF012}).

\begin{figure}[h]
\centerline{
\includegraphics[width=9cm,height=7cm]{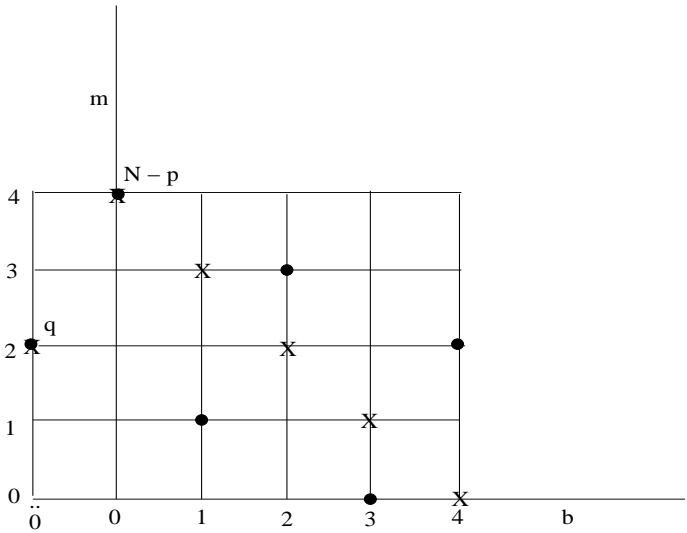}
}
\caption{
\scriptsize{Illustration of the line $m=M^{(c)}_{q,p}(b)$ in the $b-m$
plane, for the dimensionality $N=5$ and the particular pair of ``phase-space" values $q=2$
and $p=1$.
The dots correspond to the choice $c=0$ of the phase parameter and
the crosses to $c=-1/2$.
}
}
\label{line_bm_plane}
\end{figure}

Notice that our approach is based on a 
``dual affine-plane geometry" (DAPG) 
(where the points represent states), in contrast to the 
``affine-plane geometry" (APG) 
(where the lines represent states) used, e.g., in
Refs. \cite{wootters87}, and \cite{klimov_et_al_2006}.
In Ref. \cite{wootters_2006} both approaches are discussed.

i) In the present (DAPG) approach one has $N(N+1)$ points and $N^2$ lines. 
The points represent MUB projectors, and the lines, Wigner operators.
The lines defined by $m=M^{(c)}_{q,p}(b)$ depend on the freedom to choose the phase parameter $c$.
This dependence is illustrated in Fig. \ref{line_bm_plane} for the particular cases $c=0$ and $c=-1/2$, the latter having played a particular role in our earlier discussion.
%%%%%%%%%%%%%%%%%%%%%%%%%%%%%%%%%%%%%%
%%%%%%%%%%%%%%%%%%%%%%%%%%%%%%%%%%%%%%   
The matrix elements of the line operator, Eqs. (\ref{nPn'}), (\ref{nPn' 2}),
reduce, when the choice $c=-1/2$ is adopted, to the 
``standard Wigner operator" (\ref{nPn' c=-1/2}), a relatively simple expression which, for convenience, we reproduce here
%%%%%%%%%%%%%%%%%
\begin{equation}
\langle n | \hat{P}_{qp} | n' \rangle 
= \delta _{n+n',2q}
\omega^{(n-n')p}, \;\;\;\; \forall n,n'.
\label{nPn' c=-1/2 2}
\end{equation}
%%%%%%%%%%%%%%%%%
In contrast, the choice $c \neq -1/2$ leads to the general "non-standard", more complicated, Wigner operator of Eqs. (\ref{nPn'}), (\ref{nPn' 2}).
Notice, however, that the line operators are orthogonal $\forall c$, as shown in Eq. (\ref{orthogon. Pj}).

We thus see that the phase ambiguity arises via the ambiguity in defining the line equation $M_{qp}^{(c)}(b)$, giving ``standard" or ``non-standard" Wigner operators.

ii) In the (APG) approach of Ref. \cite{wootters87} one contemplates $N^2$ points and $N(N+1)$ lines.
One has a freedom in defining the point operators $\hat{A}_{\alpha}$,
$\alpha = (q,p)$, so that they are ``standard" or "non-standard" Wigner operators.
The latter are similar to the ``non-standard" Winger operators we get in the DAPG approach when $c\neq -1/2$. 
The lines give MUB projectors.

We thus see that the phase ambiguity arises via the ambiguity in defining the point operators, such that they give ``standard" or ``non-standard" Wigner operators.

In addition, a particular freedom mentioned in Ref.  \cite{wootters87}, p. 11, is a linear transformation of the phase space coordinates, leaving their basic Eqs. (A1')-(A3') intact.
 
Ref. \cite{klimov_et_al_2006} (Sec. 3.6)
%and \cite{klimov_et_al_2016} 
finds, too, a freedom in the election of a phase in the definition of Wigner's function for the discrete case.
The authors use an APG approach, so we can apply arguments similar to those used above for the APG approach of Ref. \cite{wootters87}.
In addition, Ref. \cite{klimov_et_al_2006} describes a 
phase freedom which, to the best of our knowledge, is not the same as the one found in the present paper.
The authors discuss the freedom in the choice of the ``rotation  operator" in the plane: this rotation transforms among the bases.
In our approach, the various bases are ``vertical lines" 
(see Fig. \ref{line_bm_plane}) in the b-m plane;
we have chosen for them the form given in Eq. (\ref{mub e-vectors}), which is kept fixed in all our analysis, without further changes in its structure.
Also, Ref. \cite{klimov_et_al_2006}
%{klimov_et_al_2016} 
finds a relation with parity which seems to be independent of the choice of their phase $\phi$.
In contrast, in our case, similar relations hold only for the choice $c=-1/2$.

%%%%%%%%%%%%%%%%%%%%%%%
\section{Conclusions}
\label{concl}

We considered the mapping of Quantum Mechanics from Hilbert space to phase space,
which is arguably the most intuitive form of the three Quantum Mechanics formulations: 
Hilbert space, summation over paths and phase space. 
We demonstrated that for a given finite-dimensional Hilbert space there are several phase-space forms possible which differ in what we termed a phase parameter. 
We showed that one particular value of the phase parameter brings the various expressions for the discrete case to a form similar to that for the continuous case.
In particular, our approach provides a novel view of the role of parity in such mappings: the ``most natural" choice for the phase parameter gives rise to parity, in a way that parallels the situation for the case of a Hilbert space  for continuous variables.
We briefly compared the phase freedom discussed in this paper with that found by other authors.
%%%%%%%%%%%%%%%%%%%%%%%%%%%%%%%%%%%%%%%%%%

%%%%%%%%%%%%%%%%%%%
\begin{acknowledgements}

One of the authors (PAM) acknowledges supposrt by DGAPA, under 
Contract No. IN109014.
He also acknowledges the kind hospitality of the Physics Department of the Technion, where this investigation was initiated.

\end{acknowledgements}



%%%%%%%%%%%%%%%%%%%
\begin{appendix}

\section{Schwinger operators}
\label{schwinger}

We consider an $N$-dimensional Hilbert space spanned by $N$ orthonormal states
$|q\rangle$, with $q=0,1, \cdots ,(N-1)$, and subject to the periodic condition
$|q+N\rangle=|q\rangle$;
they are designated as the ``reference basis", or ``computational basis" of the space.
We introduce the unitary operators $\hat{X}$ and $\hat{Z}$ \cite{schwinger}, which fulfill the periodic condition
%%%%%%%%%%%%%%%%%%%%%%%
$
\hat{X}^N = \hat{Z}^N =
\hat{\mathbb{I}}
$
%%%%%%%%%%%%%%%%%%%%%%%%%%%%%%
($\hat{\mathbb{I}}$ being the unit operator)
and are defined by their action on the states of the reference basis as
%%%%%%%%%%%%%%%%%%%%%%%%%%%%%%%
%\begin{subequations}
\begin{eqnarray}
\hat{Z}|q\rangle
&=&\omega^q|q\rangle, \;\;\;\; \omega=e^{2 \pi i/N},
\label{Z}  \\
\hat{X}|q\rangle &=& |q+1\rangle ,
\label{X}
\end{eqnarray}
%\label{Z,X}
%\end{subequations}
%%%%%%%%%%%%%%%%%%%%%%%%%%%%%%
leading to the commutation relation
%%%%%%%%%%%%%%%%%%%%%%%%%%%%%%
\begin{equation}
\hat{Z}\hat{X}=\omega \hat{X}\hat{Z} .
\label{comm Z,X}
\end{equation}
%%%%%%%%%%%%%%%%%%%%%%%%%%%%%%
The two operators $\hat{Z}$ and $\hat{X}$ form a complete algebraic set \cite{schwinger},
so that any operator defined in our $N$-dimensional Hilbert space can be written as a function of $\hat{Z}$ and $\hat{X}$.

We introduce the Hermitean operators $\hat{p}$ and $\hat{q}$ through the equations
\cite{wootters87,revzen-WF012,mello-revzen-PRA2014,durt_et_al,de_la_torre-goyeneche}
%%%%%%%%%%%%%%%%%%%%%%%%%%%%%%%
%\numparts
\begin{eqnarray}
\hat{X}
&=& \omega^{-\hat{p}}
= e^{-\frac{2\pi i}{N}\hat{p}} \; ,
\label{X(p)} \\
\hat{Z}
&=& \omega^{\hat{q}}
=e^{\frac{2\pi i}{N}\hat{q}} \; .
\label{Z(q)}
\end{eqnarray}
%\label{X(p),Z(q)}
%\endnumparts
%%%%%%%%%%%%%%%%%%%%%%%%%%%%%%
As {\em $\hat{X}$ performs translations in the variable $q$} and 
{\em $\hat{Z}$ in the variable $p$}, we designate $\hat{p}$ and $\hat{q}$ as {\em ``momentum-like"} and 
{\em ``position-like"} operators, respectively.

What we defined as the reference basis can thus be considered as the ``position basis".
With (\ref{comm Z,X}) and definitions (\ref{X(p)}), (\ref{Z(q)}), the commutator of $\hat{q}$ and $\hat{p}$ reduces, in the continuous limit 
\cite{durt_et_al,santhanam,de_la_torre-goyeneche},
to the standard one, $[\hat{q},\hat{p}]=i$.
%%%%%%%%%%%%%%%%%%%%%%%%%%%%%%

%%%%%%%%%%%%%%%%%%%%%%%%%%%%%%
\section{MUB}
\label{mub}

The operators $\hat{X}\hat{Z}^b$, $b=0, \cdots N-1$ define $N$ of the $N+1$ MUB, [see Eqs. (\ref{e-value eqn mub}), (\ref{mub e-vectors}) below], while the operator $\hat{Z}$ defines the reference basis.
The operator $\hat{X} \hat{Z}^{b}$ possesses $N$ eigenvectors, denoted by
$|m,b\rangle$ (see Ref. \cite{bandyopadhyay_et_al_2001_2002} and
Eqs. (10), (11) of Ref. \cite{revzen-WF012})
%%%%%%%%%%%%%%%%%%%%%%%%%%%%%%
%\numparts
\begin{eqnarray}
\hat{X}\hat{Z}^{b} |m,b\rangle
&=& \omega^m |m;b\rangle ; \;\;\;\;\; b,m = 0,1, \cdots , N-1 \; ,
\label{e-value eqn mub} \\
|m;b\rangle
&=& \frac{1}{\sqrt{N}}\sum_{q=0}^{N-1}\omega^{\frac{b}{2}q(q-1)-qm} |q\rangle .
\label{mub e-vectors}
\end{eqnarray}
%\label{MUB}
%\endnumparts
%%%%%%%%%%%%%%%%%%%%%%%%%%%%%%
Here, $|q\rangle$ ($q=0, \cdots, N-1$) denote the $N$ states of the reference basis.
The states with $b=0$ are eigenstates of $\hat{p}$
%%%%%%%%%%%%%%%%%%%%%%%%%%%%%%
%\numparts
\begin{eqnarray}
|m;0\rangle
%&=& \frac{1}{\sqrt{N}}\sum_{n=0}^{N-1}
%e^{-\frac{2\pi i}{N}mq} |q\rangle,
%\label{b=0 states a} \\
&=&\frac{1}{\sqrt{N}}\sum_{q=0}^{N-1}
e^{\frac{2\pi i}{N}(N-m)q} |q\rangle \; ,
\label{b=0 states b}\\
&=& |p=-m=(N-m) {\rm Mod}[N]\rangle.
\label{b=0 states c}
\end{eqnarray}
%\endnumparts
%%%%%%%%%%%%%%%%%%%%%%%%%%%%%%
%%%%%%%%%%%%%%%%%%%%%%
%\numparts
%\begin{eqnarray}
%W_{\hat{A}}(q,p)
%&=&\frac{1}{N}
%\left\{
%\sum_{b=0}^{N-1}\sum_{k=1}^{N-1}
%\tilde{W}_{\hat{A}}(k,b)
%e^{i\frac{2\pi}{N}k(-p + b q)}
%+ \sum_{l=0}^{N-1}
%\tilde{W}_{\hat{A}}(l) e^{i\frac{2\pi}{N}lq }
%\right\} ,
%\label{discreteWF a}
%\nonumber \\ \\
%\tilde{W}_{\hat{A}}(k,b)
%&=& {\rm Tr}\left\{\hat{A}\left[\left(\hat{X}\hat{Z}^b \right)^k %\right]^{\dagger} \right\}     \; ,
%\label{discreteWF b}\\
%\tilde{W}_{\hat{A}}(l)
%&=& {\rm Tr}\left[\hat{A} \left(\hat{Z}^l\right)^{\dagger} \right] \; .
%\label{discreteWF c}
%\end{eqnarray}
%\label{discreteWF}
%\endnumparts
%%%%%%%%%%%%%%%%%%%%%%

\section{Orthogonality of the operators $\hat{P}^{(c)}_{qp}$, 
Eq. (\ref{orthogon. Pj})}
\label{orthogonality of P's}

We show the orthogonality, as given by Eq. (\ref{orthogon. Pj}), of the operators $\hat{P}^{(c)}_{q,p}$ defined in Eq. (\ref{P(q,p)}).
We first note the following two features of the lines described in the text.

1. We assume the phase $c$ to be fixed;
then a line is defined by the pair of numbers $q,p$. 
Lines $\hat{P}^{(c)}_{q,p}, \hat{P}^{(c)}_{q',p'}$ may be either identical, which means that $p=p'$ and $q=q'$, 
or distinct, meaning that either 

i) $q \neq q'$, $p=p'$, or 

ii) $q=q'$, $p\neq p'$, or 

iii) $q\neq q'$ and $p \neq p'$. 

If two lines are distinct, then they have one point in common, i.e. there exists (one) $b$ wherein the two lines intersect. 
Every line has {\em one} common "point" with every other line. 
If two lines have two points in common, they are identical.

Let line 1 be given by $(q,p)$ and line 2 by $(q',p')$. We have the possibilities: \\
i, iii) The lines differ if $q\ne q'$. 
Here both cases $p=p'$ or $p\ne p'$ imply $b=\frac{p-p'}{q-q'}$ 
($b=0$ for the first case) - a unique value in either case. \\
ii) If the lines differ via  $p\ne p'$ but $q=q'$, the common point is at $b=\ddot{0}$. 

The uniqueness of the solutions imply that distinct lines cannot have more than one point in common.

2. All the ``points" are MUB projectors , i.e. 
%%%%%%%%%%%%%%%%%%%
\begin{equation}
{\Bbb{P}}^{(c)}(q,p;b)=\left|M^{(c)}_{q,p}(b); b \right\rangle 
\left\langle M^{(c)}_{q,p}(b); b\right| \; .
\label{mub projectors}
\end{equation}
%%%%%%%%%%%%%%%%%%%
For $b\ne b'$ (including $b=\ddot{0}$) we have
%%%%%%%%%%%%%%%%%%%
\begin{equation} 
{\rm Tr} \left[ {\Bbb{P}}^{(c)}(q,p;b){\Bbb{P}}^{(c)}(q,p;b') \right]
=\frac{1}{N} \; . 
\label{}
\end{equation}
%%%%%%%%%%%%%%%%%%%

The actual proof of orthogonality involves computing the LHS of 
Eq. (\ref{orthogon. Pj}), giving
%%%%%%%%%%%%%%%%%%%%%%%%%%%%%%%%%%%%%
%\begin{subequations}
\begin{eqnarray}
{\rm Tr} \left[ \hat{P}^{(c)}_{q,p}\hat{P}^{(c)}_{q',p'} \right]
&=&\sum_{b=\ddot{0}}^{N-1}{\rm Tr}\left[\Bbb{P}^{(c)}(q,p;b) 
\Bbb{P}^{(c)}(q',p';b)\right]
+\sum_{b\ne b'}{\rm Tr}\left[\Bbb{P}^{(c)}(q,p;b)\Bbb{P}^{(c)}(q',p';b') \right]
\nonumber \\
&& 
-\sum_{b=\ddot{0}}^{N-1}{\rm Tr}\left[\Bbb{P}^{(c)}(q,p;b)\right]
-\sum_{b=\ddot{0}}^{N-1}{\rm Tr}\left[\Bbb{P}^{(c)}(q',p';b)\right]
+{\rm Tr}\; \Bbb{I}, \\
&\equiv& A+B+C+D+E  \; .
\end{eqnarray}
%\end{subequations}
%%%%%%%%%%%%%%%%%%%%%%%%%%%%%%%%%%%%%

\underline{Calculation of $A$}.
The first term, A, involves a sum over $N+1$ traces of the product of two projectors, both in the same basis $b$.

i) If $q=q', p=p'$, the lines are identical: then the sum reduces to $N+1$ traces of projectors, each giving 1, the result thus being $N+1$. 

ii) If the lines are distinct, they have only one common point at $b$; 
we thus have the trace of a projector, giving 1. 
All other $N$ terms involve the trace of a product of two orthogonal projectors and do not contribute. Thus
%%%%%%%%%%%%%%%%%%  
\begin{equation}
A=\sum_b Tr \Bbb{P}^{(c)}(q,p;b)\Bbb{P}^{(c)}(q',p';b)
= (N+1) \delta_{qq'}\delta_{pp'}
+1\cdot(1- \delta_{qq'}\delta_{pp'})
= 1+ N \delta_{qq'}\delta_{pp'} \; .
%\begin{cases}1\;\;(q,p)\ne(q',p')\\
%N+1 \;(q,p)=(q',p')\end{cases}.
\end{equation}
%%%%%%%%%%%%%%%%%%
\underline{Calculation of $B$.}
The second term involves traces of projectors of distinct bases. 
Since these are MUB projectors, each term gives $1/N$. 
There are $N(N+1)$ terms in the sum, giving $N(N+1)/N=N+1$:
%%%%%%%%%%%%%%%%%%
\begin{equation}  
B=\sum_{b\ne b'}Tr \Bbb{P}^{(c)}(q,p;b)\Bbb{P}^{(c)}(q',p';b')=N+1.
\end{equation}
%%%%%%%%%%%%%%%%%%
\underline{Calculation of $C$ and $D$.}
The third and fourth terms, C and D, involve traces of projectors each multiplied by unity. There are $N+1$ terms, each giving 1. 
We thus find
%%%%%%%%%%%%%%%%%%
\begin{equation}
C=-\sum_b {\rm Tr} \Bbb{P}^{(c)}(q,p;b)
= D=-\sum_b Tr \Bbb{P}^{(c)}(q',p';b')
=-(N+1).
\end{equation}
%%%%%%%%%%%%%%%%%%
\underline{Calculation of $E$.}
The last term, E, involves the trace of unity, giving $N$.

Adding up the five terms, we finally obtain:
%%%%%%%%%%%%%%%%%%
\begin{equation}
{\rm Tr} \left[\hat{P}^{(c)}_{q,p}\hat{P}^{(c)}_{q',p'}\right]
=N\delta_{q,q'}\delta_{p,p'}\; .
\end{equation}
%%%%%%%%%%%%%%%%%%

%%%%%%%%%%%%%%%%%%%%%%%%%%%%%%%
\section{Derivation of Eq. (\ref{WF continuous 3})}
\label{alternative structure for continuous case}

The definition of WWT, Eq. (\ref{WF continuous a}), is based on the notion that the Weyl operators \cite{weyl}
%%%%%%%%%%%%%%%%%%%%%%%%%%%%%%%%%%%%%%%%%%%
%\numparts
\begin{eqnarray}
\mathbb{U}
={\rm e}^{i(\beta \hat{p} +\alpha \hat{q})}
\label{Ucont}
\end{eqnarray}
%%%%%%%%%%%%%%%%%%%%%%%%%%%%%%%%%%%%%%%%%%%
form a complete and orthogonal operator basis
\cite{leonhardt,schleich,khanna-mello-revzen}.
For the purpose of comparing the continuous and discrete cases
in Sec. \ref{WWT cont variables},
we introduce the alternative set of complete and orthogonal operators
%%%%%%%%%%%%%%%%%%%%%%%%%%%%%%%%%%%%%%%%%%%
\begin{eqnarray}
\mathbb{V}
= {\rm e}^{i\beta \hat{p}} \; {\rm e}^{i\alpha \hat{q}} \; ,
\label{Zcont}
\end{eqnarray}
\label{Ucont,Zcont}
%\endnumparts
%%%%%%%%%%%%%%%%%%%%%%%%%%%%%%%%%%%%%%%%%%%
and express the WWT in terms of them.

Making the change of variables $v'=-v$ and relabelling $v'$ again as $v$, we obtain, from Eq. (\ref{WF continuous a})
%%%%%%%%%%%%%%%%%%%%%%%%%%%%%%%
\begin{equation}
W_{\hat{A}}(q,p)
= \frac{1}{2\pi} {\rm Tr}\int_{-\infty}^{\infty}
\int_{-\infty}^{\infty}
\hat{A} {\rm e}^{-i(u\hat{q} - v\hat{p})} {\rm e}^{i(uq - vp)} du dv \; .
\label{WF continuous 1}
\end{equation}
%%%%%%%%%%%%%%%%%%%%%%%%%%%%%%%
We use polar coordinates
%%%%%%%%%%%%%
\begin{equation}
u = r \cos \theta \; , \;\;\;  v = r \sin \theta
\label{polar}
\end{equation}
%%%%%%%%%%%%%
and write
%%%%%%%%%%%%%%%%%
%\numparts
\begin{eqnarray}
&& W_{\hat{A}}(q,p)
= {\rm Tr}\int_{0}^{\infty} r dr
\int_{0}^{2 \pi} d \theta \;
\hat{A} {\rm e}^{-i r \sin \theta (- \hat{p} + \cot \theta \; \hat{q} )}
\nonumber  \\
&& \times  {\rm e}^{i r \sin \theta (- p + \cot \theta \; q)}
\label{WF continuous 2 a}  \\
&&  =
{\rm Tr}\int_{0}^{\infty} r dr
\int_{0}^{2 \pi} d \theta \;
\hat{A} \left[{\rm e}^{i(-\hat{p} + \cot \theta \; \hat{q})} \right]^{-r \sin \theta}  \nonumber \\
&& \hspace{1cm} \times {\rm e}^{i r \sin \theta (- p + \cot \theta \; q)} \; .
\label{WF continuous 2 b}
\end{eqnarray}
\label{WF continuous 2}
%\endnumparts
%%%%%%%%%%%%%%%%%%%%%%%%%%%%%%%
Recalling the BCH identity
\begin{equation}
{\rm e}^{i(\alpha \hat{q} + \beta \hat {p})}
= {\rm e}^{-\frac{i}{2}\alpha \beta} {\rm e}^{i \beta \hat {p}}
{\rm e}^{i \alpha \hat {q}} \; ,
\label{BCH}
\end{equation}
%%%%%%%%%%%%%
and choosing $\beta = -1$, $\alpha = \cot \theta$, we can write
%%%%%%%%%%%%%
%\numparts
\begin{eqnarray}
{\rm e}^{i(-\hat{p} + \cot \theta \; \hat{q})}
&=& {\rm e}^{\frac{i}{2}\cot \theta}
{\rm e}^{-i \hat{p}} \;{\rm e}^{i \cot \theta \; \hat{q}} \; ,
\label{BCH 1 a} \\
\left[{\rm e}^{i(-\hat{p} + \cot \theta \; \hat{q})}\right]^{-r \sin \theta}
&=& \left( {\rm e}^{-i \hat{p}} \;{\rm e}^{i \cot \theta \; \hat{q}}\right)
^{-r \sin \theta}
{\rm e}^{-\frac{i}{2}\cot \theta \; r \sin \theta} \; ,
\nonumber  \\
\label{BCH 1 b}
\end{eqnarray}
\label{BCH 1}
%\endnumparts
%%%%%%%%%%%%%
so (\ref{WF continuous 2 b}) takes the form of Eq. (\ref{WF continuous 3}) given in the text.

\end{appendix}

%\begin{figure}
%  \includegraphics{example.eps}
%\end{figure}
%\begin{figure*}
%  \includegraphics[width=0.75\textwidth]{example.eps}
%\end{figure*}
%\begin{table}[h]
%\begin{tabular}{lll}
%\hline\noalign{\smallskip}
%irst & second & third  \\
%\end{tabular}
%\end{table}
%%%%%%%%%%%%%%%%%%%

%%%%%%%%%%%%%%%%%%%

%%%%%%%%%%%%%%%%%%%

%%%%%%%%%%%

\end{document}